\documentstyle[prl,twocolumn,aps,epsf]{revtex}
\input{psfig}

\def\Hqb{{\cal H}_{\rm qb}}
\def\Hint{{\cal H}_{\rm int}}
\def\Eqb{E_{\rm qb}}
\def\Eint{E_{\rm int}}
\def\Epar{E_{\rm int}^{\parallel}}
\def\Eper{E_{\rm int}^{\perp}}
\def\EJ{E_{\rm J}}
\def\Eset{E_{\rm SET}}

\def\Gmix{\Gamma_{\rm mix}}
\def\Vtr{V_{\rm tr}}

\def\I{{\bar I}}
\def\initcond{\gamma_0}
\newcommand\G[1]{\Gamma^{{#1}}}
\newcommand\DM[2]{\rho_{#1}^{#2}}      %  #1=N  #2=ij
\newcommand\HDM[1]{\hat\rho_{#1}}      %  #1=N
\def\i{{\rm i}}

\begin{document}

\title{Statistics and noise in a quantum measurement process}

\author{Yuriy Makhlin$^{1,2}$, Gerd Sch\"on$^{1,3}$, and Alexander 
Shnirman$^{1}$}

\address{
$^1$Institut f\"ur Theoretische Festk\"orperphysik,
Universit\"at Karlsruhe, D-76128 Karlsruhe, Germany\\
$^2$Landau Institute for Theoretical Physics, 
Kosygin st. 2, 117940 Moscow, Russia\\
$^3$Forschungszentrum Karlsruhe, Institut f\"ur Nanotechnologie,
D-76021 Karlsruhe, Germany}

\maketitle

\begin{abstract}
The quantum measurement process by a single-electron transistor or a quantum
point contact coupled to a quantum bit is studied.  We find a unified
description of the statistics of the monitored quantity, the current, in the
regime of strong measurement and expect this description to apply for a wide 
class of quantum measurements.  We derive the probability distributions for the
current and charge in different stages of the process.  In the parameter regime
of the strong measurement the current develops a telegraph-noise behavior which
can be detected in the noise spectrum.
\end{abstract}

\paragraph*{Introduction.}
The long-standing interest in fundamental questions of the quantum
measurement received new impetus by the experimental progress in
mesoscopic physics and increasing activities in quantum state
engineering. The basic idea is to use as meter a device, able
to carry a macroscopic current, which is
coupled to the quantum system such that the conductance
depends on the quantum state. By monitoring the current one
performs a quantum measurement, which,
in turn, causes a dephasing of the quantum
system~\cite{Aleiner,Levinson,Gurvitz}. The dephasing has
been demonstrated in an experiment of Buks et al.~\cite{Buks} 
where a quantum dot is embedded in one arm of a `which-path'
interferometer. The current through a quantum point contact (QPC)
in close proximity to the dot suppresses the interference.
However, since passing electrons interact with the current only
for a short dwell-time in the dot, the meter fails to 
distinguish between two possible paths of individual electron;
only a reduction of interference has been observed. 
This situation is referred to as a {\em weak} measurement.

For a {\em strong} quantum measurement a different setting is needed,
where a closed quantum system is observed by a meter.  
Then a sufficiently long observation may 
provide information about the quantum state.  This situation is
realized when a single-electron transistor (SET) is coupled to a Josephson
junction single-charge box, which for suitable parameters serves as a quantum
bit (qubit)~\cite{Our_PRL,Our_Nature}.  The analysis of the time evolution of
the density matrix of the coupled system demonstrates the mutual influence
between qubit and meter, i.e.\ measurement and dephasing~\cite{Our_PRB}.

The measurement process is characterized by three time scales. 
On the shortest, the dephasing time $\tau_\varphi$, the phase coherence 
between two eigenstates of the qubit is lost, while their occupation 
probabilities remain unchanged.  Later measurement-induced transitions
mix the eigenstates,  changing
their occupation probabilities on a time scale $t_{\rm mix}>\tau_\varphi$ and
erasing information about the initial state of the qubit.  The origin
of the mixing is that the charge operator (the
measured quantity) and the qubit's Hamiltonian do not commute.
The third time scale appears in the dynamics of the current in the SET.
Consider the probability distribution $P(m,t)$ that $m$ 
electrons have tunneled through the SET by time $t$.
It was shown~\cite{Our_PRB} that after a certain time $t_{\rm meas}$
information about the 
state of the qubit can be extracted by reading out $m$.
As expected from the basic principles of quantum mechanics
the observation of the qubit first of all disturbs its state. 
Hence $t_{\rm meas} \ge \tau_\varphi$.  The measurement process is only
effective if the mixing is 
slow, $t_{\rm mix}\gg t_{\rm meas}$.
In the opposite limit of strong mixing the information about the qubit's 
state is lost before a read-out is achieved.

The distribution function $P(m,t)$ describes the statistics of the charge which
has tunneled.  The distribution of the current in the
SET $p(I,t)$  and current-current correlations  require, furthermore, the
knowledge of correlations of the values of $m$ 
at different times. In earlier papers
on the statistics in a SET~\cite{Our_PRB} or a 
QPC~\cite{Gurvitz_Measurement,Stodolsky} the
behavior of $P(m,t)$ at times shorter than $t_{\rm mix}$ was derived. 
Effects due to the additional knowledge acquired by an observer~\cite{Korotkov}, 
and the possible influence of the wave-function collapse on the monitored 
current~\cite{Gurvitz_Measurement} were also discussed.
Here we develop a systematic approach, based on the time evolution 
(Schr\"odinger equation) of the
density matrix of the coupled system.  This approach allows us to
study averages and correlators of the current and charge.  Since, due
to shot noise, instantaneous values of the current fluctuate 
strongly, we study the current $\I$, averaged over a
finite time interval $\Delta t$.  We calculate the mixing time in a SET and 
derive analytic expressions for $p(\I,{\Delta t},t)$, as well as $P(m,t)$, valid 
on both short and long time scales.
We study the noise spectrum of the current
and find that in the limit of strong measurement
($t_{\rm meas}<t_{\rm mix}$) the long-time dynamics is characterized by 
{\em telegraph noise}, with jumps between two possible current values, 
corresponding to two qubit's eigenstates.

The results are  of immediate experimental interest.  
A recent experiment demonstrated  the quantum coherence in a macroscopic
superconducting electron box~\cite{Nakamura_Nature},
but the coherence 
time was limited by the measuring device.  The SET-based measurement should
extend the coherence time, which combined with experimental progress
in fast measurement techniques~\cite{Schoelkopf} should increase the
maximum number of coherent quantum manipulations.

\begin{figure}  
\centerline{\hbox{\psfig{figure=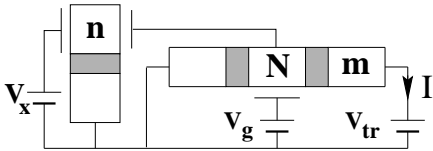,width=0.66\columnwidth}}}
\caption[]{\label{Figure:Qubit+SET}
The circuit of a qubit and a SET
used as a meter.}
\end{figure}

%%%%%%%%%%%%%%%%%%%%%%%%%%%%%%%%%%%%%%%
{\it Master equation for the measurement by a SET.}
The system of a qubit coupled to a SET is shown in 
Fig.~\ref{Figure:Qubit+SET}. The qubit is a Josephson junction
in the Coulomb blockade regime. Its dynamics is limited to a two-dimensional
Hilbert space spanned by two charge states, with $n=0$ or $1$ extra Cooper
pair on a superconducting island.  The island is coupled capacitively to the
middle island of the SET, influencing the transport current.  The SET
is kept in the off-state during manipulations on
the qubit~\cite{Our_PRB}, with no dissipative current and no additional
decoherence.  To perform the measurement, the
transport voltage is switched to a sufficiently high value, so that the current
starts to flow in the SET.

%%%%%%%%%%%%%%%%%%%%%%%%%%%%%%%%%%%%%%%%%%%%%%%%%%%%%%
%%%%%%%%%%%%%%%%%%%%%%%%%%%%%%%%%%%%%%%%%%%%%%%%%%%%%%

The Hamiltonian of the system is given by~\cite{Our_Book}
\begin{equation}
{\cal H}={\cal H}_{\rm SET} +\Hqb^N
+{\cal H}_{\rm T} +{\cal H}_{\rm \psi}
\; .
\label{H}
\end{equation}
The first term is the charging energy of the transistor, quadratic in the charge 
$Ne$ on the middle island, ${\cal H}_{\rm SET}=\Eset(N-N_0)^2$. The 
induced charge $N_0 e$ is determined by the gate voltage $V_{\rm g}$
and other voltages in the circuit.
The Hamiltonian of the qubit, $\Hqb^N=\Hqb+N\Hint$, includes 
the Hamiltonian of the uncoupled qubit $\Hqb$ and the Coulomb interaction with 
the SET $N\Hint$
(split into two terms for later convenience.)
In the basis of the qubit's charge states they are given 
by $\Hqb=\frac{1}{2}(\Eqb\hat\sigma_z-\EJ\hat\sigma_x)$
and $N\Hint=\Eint N\hat\sigma_z$. The charging energy scales 
$\Eset$, $\Eqb$ and $\Eint$ are determined by capacitances in 
the circuit, while $\EJ$ is the Josephson coupling. 
We consider the eigenstates
of $\Hqb$, $|0\rangle$ and $|1\rangle$, as the logical states
of the qubit. In this basis, 
$\Hqb$ is diagonal, with the level spacing
$\Delta E \equiv (\EJ^2+\Eqb^2)^{1/2}$, while 
$\Hint=\left(\begin{array}{cc}\Epar&\Eper\\\Eper&-\Epar\end{array}\right)$,
where $\Epar \equiv \Eint\Eqb/\Delta E$ and $\Eper \equiv \Eint\EJ/\Delta E$.
The term ${\cal H}_{\rm \psi}$ describes the Fermions in the
island and electrodes of the SET, while ${\cal H}_{\rm T}$ governs
the tunneling in the SET.
Here we assume weak coupling to the environment, with relaxation slower than the 
SET-induced mixing. The opposite limit is dicussed in Ref.~\cite{Our_Book}.

The full density matrix can be reduced to 
$\rho^{iN\phantom{'}m\phantom{'}}_{jN'm'}(t)$ by tracing over microscopic 
degrees of freedom and keeping track only of $N$ and $m$, the number of 
electrons which have tunneled through the SET. 
Here $i,j=0,1$ refer to a qubit's basis. A closed set of equations can be 
derived for $\DM{N}{ij}(m)\equiv\rho^{iNm}_{jNm}$, the diagonal 
entries of the density matrix in $N$ and $m$~\cite{Schoeller_PRB}.
Solving these equations, we analyze the evolution of the reduced 
density matrix of the qubit, 
$\varrho^{ij}(t) \equiv \sum_{N,m}\DM{N}{ij}(m,t)$, 
as well as $P(m,t) \equiv \sum_{N,j}\DM{N}{jj}(m,t)$ and other statistical 
characteristics of the current in the SET.

At low temperatures and transport voltages only two charge 
states of the middle island of the SET, with $N$ and $N+1$ electrons, 
contribute to the dynamics. 
Translational invariance in $m$-space suggests the Fourier transformation 
$\DM{N}{ij}(k)\equiv \sum_m e^{-\i km} \DM{N}{ij}(m)$. Expanding in the 
tunneling term to lowest order, we obtain the following 
master equation (cf. Refs.~\cite{Our_PRB,Our_Book}):
\begin{eqnarray}
\hbar\frac{d}{dt}\left(\begin{array}{c}\HDM{N\phantom{+1}}\\
\HDM{N+1}\end{array}\right)
&+&\left(\begin{array}{c}
\i\,[\Hqb,\HDM{N}]\\[3pt]
\i\,[\Hqb+\Hint,\HDM{N+1}]
\end{array}\right)
\nonumber
\\
&=&
\left(\begin{array}{cc}
-\check\Gamma_L & e^{\i k}\check\Gamma_R\\
\phantom{-}\check\Gamma_L & -\check\Gamma_R
\end{array}\right)
\left(\begin{array}{c}\HDM{N\phantom{+1}}\\
\HDM{N+1}\end{array}\right)
\; .
\label{MasterEq}
\end{eqnarray}
Here the operators
\begin{eqnarray}
\check\Gamma_L\hat\rho&\equiv&
\Gamma_L\hat\rho
+\pi\alpha_L\;[\Hint,\hat\rho]_+
\; ,
\label{GL}
\\
\check\Gamma_R\hat\rho&\equiv&
\Gamma_R\hat\rho
-\pi\alpha_R\;[\Hint,\hat\rho]_+
\; .
\label{GR}
\end{eqnarray}
represent the tunneling rates in the left and right junctions, with
$\alpha_{\rm L/R}\equiv R_{\rm K}/(8\pi^3R^{\rm T}_{\rm L/R})$
being the tunnel conductance of the junctions in units of
the resistance quantum $R_{\rm K}=h/e^2$.
The rates are fixed by the potentials
$\mu_L$, $\mu_R=\mu_L+\Vtr$ of the leads:
$\Gamma_L=2\pi\alpha_L[\mu_L-(1-2N_0)\Eset]$ and $\Gamma_R=
2\pi\alpha_R[(1-2N_0)\Eset-\mu_R]$.
They define the tunneling rate 
$\Gamma=\Gamma_L\Gamma_R/(\Gamma_L+\Gamma_R)$ through the SET.
The last terms in Eqs.~(\ref{GL},\ref{GR}) make these rates 
sensitive to the qubit's state.

The initial condition at the beginning of the measurement,
written down in the logical basis,
\begin{equation}
\label{INITIAL_CONDITIONS}
\rho^{ij}_{N}(m,t=0) =
  \left(
  \begin{array}{cc}
  |a|^2  & ab^{*} \\
  a^{*}b & |b|^2
  \end{array}
  \right)
w_N
\;
\delta_{m0}
\ ,
\end{equation}
describes the qubit in a pure state $a|0\rangle + b|1\rangle$
and the SET in the zero-voltage equilibrium state.
One can assume that $w_N=1$ and $w_{N+1}=0$.

%%%%%%%%%%%%%%%%%%%%%%%%%%%%%%%%%%%%%%%%%%%%%%%%%%%%%%%%%%%
\paragraph*{Reduction of the master equation.}
In general the dynamics of the qubit's density matrix $\hat\varrho$,
described by the  
master equation (\ref{MasterEq}), is complicated since 
{\em dephasing} (decay of the off-diagonal entries)
and {\em mixing} (relaxation of the diagonal to their stationary values) 
may occur on similar time scales, $\tau_\varphi\approx t_{\rm mix}$.
However, under suitable conditions the mixing is slow, which is the
prerequisite for a measurement process. This is the case,
if the qubit operates in the regime with dominant
charging energy:
\begin{equation}
\label{APP_COND}
|E_{\rm qb}|, |E_{\rm qb} + 2 E_{\rm int}| \gg E_{\rm J}
\ .
\end{equation} 
Weak ($\Eint\ll\Delta E$) or strong ($\Eint\!\sim\!\Delta E$) coupling to 
the SET can be considered.
In the latter case a faster measurement is achieved (see Eq.~(\ref{tmeas})).

We first analyze the dynamics without mixing, i.e.
we put  $\Eper = 0$. In this case the time evolutions of 
$\DM{N}{ij}$ for the four  different pairs of indices $ij$  
are decoupled from each other, each being characterized by 
two eigenmodes.
The absence of mixing, further, implies the 
conservation of occupations of the logical states
$\varrho^{ii}$, for $i=0,1$,
and we find two `Goldstone' modes for $k\ll1$, with eigenvalues
\begin{equation}
\label{lambda_ii}
\lambda^{ii}_{+}(k)\approx \i\,\G{i} k-\frac{1}{2}f^{i}\G{i} k^2
\ .
\end{equation}
Here $\Gamma^{i} \equiv 
\Gamma_L^{i}\Gamma_R^{i}/(\Gamma_L^{i}+\Gamma_R^{i})$
are the tunneling rates through the SET,
$\Gamma_{L}^{0/1}=\Gamma_{L} \pm \pi\alpha_{L}\Epar$ and 
$\Gamma_{R}^{0/1}=\Gamma_{R} \mp \pi\alpha_{R}\Epar$
are the tunneling rates in the junctions for two logical states
(cf.~Eqs.~(\ref{GL},\ref{GR})), and
$f^{i} \equiv 
%[(\Gamma_L^{i})^2+(\Gamma_R^{i})^2]/[\Gamma_L^{i}+\Gamma_R^{i}]^2
1-2\G{i}/(\Gamma_L^i+\Gamma_R^i)
$
are the Fano factors responsible for the shot noise reduction. 
The other two eigenmodes decay fast, with the rates 
$\lambda^{ii}_{-} \approx -(\G{i}_L+\G{i}_R)$. 

The analysis of the eigenvalues, $\lambda^{01}_\pm(k)=[\lambda^{10}_\pm(-k)]^*$, 
of the four off-diagonal modes in $ij$, reveals the dephasing 
time $\tau_\varphi$ of the qubit by the measurement, i.e., the decay 
time of $\varrho^{01}=\DM{N}{01}(k=0)+\DM{N+1}{01}(k=0)$. It is given by  
$\tau_\varphi^{-1}={4\Gamma\Eint^{\parallel\;2}}/{(\Gamma_L+\Gamma_R)^2}$ 
if $\Epar \ll \Gamma_L+\Gamma_R$, and $\tau_\varphi^{-1}=w_N\Gamma_L + 
w_{N+1}\Gamma_R$ 
in the opposite case. 

The picture is modified by the mixing at finite $\Eper$.
We find that the mixing may be treated perturbatively if 
$|\lambda^{01}_{\pm}| \gg \Eper$, which turns to be 
the case if the condition (\ref{APP_COND}) holds. 
Then, in the second order, the degeneracy 
between the long-living modes (\ref{lambda_ii}) is lifted
and the long-time evolution of the occupations 
$\rho^{ii}(k)=\rho^{ii}_N(k)+\rho^{ii}_{N+1}(k)$
is given by a reduced master equation,
\begin{eqnarray}
&
\label{REDUCED_MASTER_EQUATION}
\displaystyle\frac{d}{dt}\left(\begin{array}{cc} \rho^{00}(k) \\
\rho^{11}(k) \end{array} \right)
=
M_{\rm red}
\left(\begin{array}{cc} \rho^{00}(k) \\
\rho^{11}(k) \end{array} \right)
\;,
&
\\[2mm]
&
\label{M_RED}
M_{\rm red} = 
\left(\begin{array}{cc}
\lambda^{00}_{+}(k) & 0
\\
0 & \lambda^{11}_{+}(k) 
\end{array}
\right) 
+
\displaystyle\frac{1}{2}\Gmix
\left(\begin{array}{cc}
-1 & \phantom{+}1
\\
\phantom{+}1 & -1
\end{array}
\right)
\ .
& 
\end{eqnarray}
For the mixing rate, $\Gmix$, we obtain
\begin{eqnarray}
\label{GMIX}
&&\Gmix \approx
\nonumber \\
&&\frac{4\Gamma \EJ^2 \Eint^2}{\Delta E^2(\Delta E+2\Epar)^2
+[\Gamma_R\Delta E+\Gamma_L(\Delta E+2\Epar)]^2}
\;.
\end{eqnarray}

To understand the role of the mixing we assume first $\Gmix = 0$ in 
Eqs.~(\ref{REDUCED_MASTER_EQUATION},\ref{M_RED}). Then, for the 
initial condition (\ref{INITIAL_CONDITIONS})
we obtain  
$\rho^{00}(k)=|a|^2 e^{\lambda_{+}^{00}(k)\;t}$,  
$\rho^{11}(k)=|b|^2 e^{\lambda_{+}^{11}(k)\;t}$,
and $P(k,t) \equiv \sum_{m} P(m,t)\,e^{-ikm}=\rho^{00}(k)+\rho^{11}(k)$. 
From this we obtain the distribution $P(m,t)$, which evolves from a  peak 
$\delta(m)$ at $t=0$ into two peaks with weights $|a|^{2}$ 
and $|b|^{2}$,  moving in $m$-space with velocities 
$\G{0}$ and $\G{1}$, and with widths growing as $\sqrt{2f^{i}\G{i}t}$. 
The peaks separate after a time
\begin{equation}
t_{\rm meas}=
\left(
	\frac{\sqrt{2f^{0}\G{0}}+\sqrt{2f^{1}\G{1}}}{\G{0}-\G{1}}
\right)^2
\;.
\label{tmeas}
\end{equation}
Thus measuring the charge $m$ after $t_{\rm meas}$ 
constitutes a strong quantum measurement~\cite{Our_PRB}.
However, at longer times $t>\Gmix^{-1}$ the mixing spoils this picture.
In particular, the occupations of the logical states 
relax to the equal-weight distribution:
$\varrho^{00}(t) - \varrho^{11}(t) \propto \exp(-\Gmix 
t)$. Therefore the two-peak structure appears only in the interval
between $t_{\rm meas} \le t \le \Gmix^{-1}$. The measurement can be
performed only if $t_{\rm meas} \ll \Gmix^{-1}$

The measurement takes longer than the dephasing, $t_{\rm
meas}\gg\tau_\varphi$.  Such measurement can be called
non-efficient~\cite{Korotkov}: the information about the qubit is contained in
the SET already after $\tau_\varphi$, but can be read out from the current only
later.

%%%%%%%%%%%%%%%%%%%%%%%%%%%%%%%%%%%%%%%%%%%%%%
{\it The quantum measurement with a QPC} can be described in a similar way.
The Coulomb interaction of the qubit with the current results in two tunneling 
rates 
$\Gamma^{0/1}=\bar\Gamma\pm\delta\Gamma/2$ for two qubit's states. 
Tracing out microscopic degrees of freedom one arrives at the master 
equation~\cite{Gurvitz} for the Fourier transform of the density matrix 
$\rho^{ij}(m)$,
which can be rewritten as
\begin{eqnarray}
\label{QPC_ME}
\hbar\frac{d}{dt}\hat\rho + \i[\Hqb,\hat\rho] &=&
\left[ \bar\Gamma\hat\rho+\frac{1}{2}\delta\Gamma[\hat\sigma_z,\hat\rho]_+
\right](e^{\i k}-1)
\nonumber\\
&&-
(4\tau_\varphi)^{-1}
e^{\i k}
[\hat\sigma_z,[\hat\sigma_z,\hat\rho]]
\;.
\end{eqnarray} 
One can show that
$\tau_\varphi^{-1} \equiv \frac{1}{2}(\sqrt{\Gamma^{0}} - \sqrt{\Gamma^{1}})^2$
is the decay rate of $\rho^{01}$. For $\rho^{ii}$ the eigenvalues are given by 
(\ref{lambda_ii}), without 
Fano factors. The measurement time (\ref{tmeas}) and the 
dephasing time coincide, implying the 100\% efficiency. 

Under the condition $E_{\rm J} \ll {\rm max}(\Delta E, \tau_\varphi^{-1})$
the perturbative treatment produces the reduced master 
equation~(\ref{REDUCED_MASTER_EQUATION},\ref{M_RED}), with
the mixing rate
\begin{equation}
\label{GMIX_QPC}
\Gmix = E_{\rm J}^2{ \tau_\varphi \over 1 + \Delta E^2\, \tau_\varphi^2}
\ .
\end{equation} 
A phenomenon, termed the Zeno (or watchdog)
effect, can be seen~\cite{Gurvitz,Gurvitz_Measurement} in the limit
$\tau_\varphi^{-1} \gg \Delta E$:  the stronger is the measurement, quantified
by $\tau_\varphi^{-1}$, the weaker is the rate of jumps between the eigenstates,
$\Gmix \approx E_{\rm J}^2\tau_\varphi$.

The analysis of the SET mixing rate (\ref{GMIX}) in terms of the Zeno
effect is 
more complicated. The rates $\tau_\varphi^{-1}$ and $\Gmix$ depend in this case 
on several parameters and no simple relation between $\Gmix$ and  
$\tau_\varphi$ is found. However, in the regime 
$E_{\rm int} \ll \Gamma_L\sim\Gamma_R \ll |\Delta E|$, these two 
rates change in opposite directions as functions of 
$\Gamma_L\sim\Gamma_R$, which is reminiscent of the Zeno physics.

%%%%%%%%%%%%%%%%%%%%%%%%%%%%%%%%%%%%%%%%%%%%%%%
{\it Statistics of charge and current.} 
The results of this section apply to the SET and QPC alike.
The statistical quantities studied depend on 
the initial density matrix (\ref{INITIAL_CONDITIONS}): e.g., 
$P(m,t)=P(m,t\,{\small |}\,\rho_{0})$. In the two-mode approximation %of Eqs.~
(\ref{REDUCED_MASTER_EQUATION},\ref{M_RED})
this reduces to a dependence 
on $\initcond \equiv \varrho^{00}-\varrho^{11}=|a|^2-|b|^2$. 
We solve Eq.~(\ref{REDUCED_MASTER_EQUATION}) to obtain 
$P(m,t\,{\small |}\,\initcond) = {\rm Tr_{\,qb}}[U(m,t)\rho_{0}]$,
where $U(m,t)$ is the inverse Fourier transform of
$U(k,t) \equiv \exp\;[M_{\rm red}(k)\;t]$.
If $\Gamma^{0/1}=\bar\Gamma\pm\delta\Gamma/2$ are close,
the resulting distribution is
\begin{equation}
\label{P(m,t)_FINAL}
P(m,t \,{\small|}\,\initcond)=
\sum_{\delta m}
{\tilde P}(m-{\delta m},t \,{\small|}\,\initcond)
\;
\frac{%
e^{-{\delta m}^2/2f\bar\Gamma t}
}
{\sqrt{2\pi f\bar\Gamma t}}
\;.
\end{equation}
The first term in the convolution (\ref{P(m,t)_FINAL}) contains two delta-peaks, 
corresponding to two qubit's logic states:
\begin{eqnarray}
&&{\tilde P}(m,t\,{\small|}\,\initcond)=
P_{\rm pl}\left({m-\bar \Gamma t \over \delta\Gamma t/2}\;,\;
\frac{1}{2}\Gmix t\; \Bigl|\;\initcond\right)
\nonumber \\
&&\ \ +\,
e^{-\Gmix t/2} \left[|a|^2\delta\left(m-\Gamma^{0} t\right)
+|b|^2\delta\left(m-\Gamma^{1} t\right)\right]
\;.
\end{eqnarray}
On the time scale $t_{\rm mix}\equiv\Gmix^{-1}$ the peaks disappear; instead
a plateau arises. It is described by
%%%%%%%%%%%%%%%%%%%%%%%%%%%%%%%%%%%%
\begin{eqnarray}
\label{Ppl}
&&P_{\rm pl}(x,\tau\,{\small|}\,\initcond)=e^{-\tau} \frac{\Gmix}{2\delta 
\Gamma}
\left\{
I_0\left(\tau
\sqrt{1-x^2}\right)
\right.
\nonumber \\
&&\qquad+\left.
\left(1+\initcond x\right)
I_1\left(\tau\sqrt{1-x^2}\right)/\sqrt{1-x^2}
\right\}
\; ,
\end{eqnarray}
at $|x|<1$ and $P_{\rm pl}=0$ for $|x|>1$. Here $I_0$, $I_1$ are the 
modified Bessel functions. At longer times the plateau 
transforms into a narrow peak centered around $m=\bar\Gamma t$. 

\begin{figure}  
\epsfxsize=\columnwidth
\centerline{\hbox{\epsffile{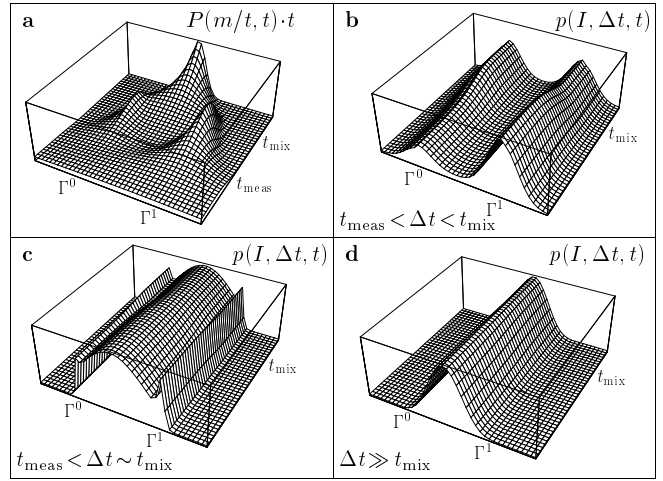}}}
\vspace{5mm}
\caption[]{\label{Figure:P(I)}
The probability distributions of the charge (a) and current
(b--d). $P(m,t)$ in (a) is rescaled, so that the peaks do not move. 
The time-axis scale is logarithmic.
}
\end{figure}

The Gaussian in Eq.~(\ref{P(m,t)_FINAL}) arises due to shot noise. Its effect is 
to smear out the distribution (see Fig.~\ref{Figure:P(I)}a).

We also calculate $P_{2}(m,t;m',t'\,{\small|}\,\rho_{0})$, the joint probability 
to have $m$ electrons at $t$ and $m'$ electrons at $t'$. This allows us to 
obtain the probability distribution
\begin{equation}
\label{P(I)_DEFINITION}
p(\I,{\Delta t},t\,{\small|}\,\rho_{0}) 
\equiv
\sum\limits_{m} P_{2}(m+\I\Delta t, t+\Delta t;m,t\,{\small|}\,\rho_{0})
\end{equation}
of the current $\I\equiv\int_t^{t+\Delta t} I(t') dt'$ averaged over the
time interval $\Delta t$. The evolution is Markovian, and we obtain:
$P_{2}(m,t;m+\Delta m,t+\Delta t) 
= {\rm Tr_{\,qb}}\,\left[
U(\Delta m,\Delta t)\,U(m,t)\rho_{0}
\right]
$ for $\Delta t > 0$. 
The derivation of $p(\I,\Delta t,t)$ thus reduces to the calculation of 
the charge distribution (\ref{P(m,t)_FINAL}) for different initial conditions:
\begin{equation}
p(\I,{\Delta t},t\,{\small|}\,\initcond)=
P(m=\I\Delta t,\Delta t\,{\small|}\,e^{-\Gmix t}\initcond)
\;.
\end{equation}
The behavior of $p(\I,\Delta t,t)$ is shown in 
Fig.~\ref{Figure:P(I)}.
A strong quantum measurement is achieved if
$t_{\rm meas} < \Delta t < t_{\rm mix}$, at times $t < t_{\rm mix}$ (see 
Fig.~\ref{Figure:P(I)}b).
In this case the measured current is close to either 
$\Gamma^{0}$ or  $\Gamma^{1}$, with 
probabilities $|a|^2$ and $|b|^2$, respectively.
At longer times a typical current pattern
is a telegraph signal jumping between $\Gamma^{0}$
and $\Gamma^{1}$ on a time scale $t_{\rm mix}$.
If $\Delta t\ll t_{\rm meas}$ the meter does not have 
enough time to extract 
the signal from the shot-noise background.
At larger $\Delta t$
the meter-induced mixing erases the information, partially ($\Delta t\sim t_{\rm 
mix}$, Fig.~\ref{Figure:P(I)}c) or completely ($\Delta t\gg t_{\rm mix}$, 
Fig.~\ref{Figure:P(I)}d), before it is read out.

The telegraph noise behavior is also seen in the current
noise. Fourier transformation of the correlator
\begin{equation}
\label{<mm>_DEFINITION}
\langle I(t)\,I(t') \rangle_{\rho_{0}}
=
\sum_{m,m'} m\,m'\,
\partial_t\partial_{t'}P_{2}(m,t;m',t'\,{\small|}\,\rho_{0})
\end{equation}
gives in the stationary case the noise spectrum,
\begin{equation}
\label{NOISE}
S_{I}(\omega)= 
2e^{2} f \bar \Gamma + 
{e^{2}\delta\Gamma^{2}\Gmix\over \omega^2 + \Gmix^{2}}
\ ,
\end{equation}
as the sum of the shot- and telegraph-noise contributions.
At low frequencies $\omega\ll\Gmix$ the latter becomes visible
on top of the shot noise
as we approach the regime of the strong measurement:
$S_{\rm telegraph}/S_{\rm shot} \approx 4t_{\rm mix}/t_{\rm meas}$.
%%%%%%%%%%%%%%%%%%%%%%%%%%%%%%%%%%%%%%%%%%%%%%%%%%%%%%
%%%%%%%%%%%%%%%%%%%%%%%%%%%%%%%%%%%%%%%%%%%%%%%%%%%%%%

To conclude, we have developed the master equation approach to 
study the statistics of currents in a SET or a QPC as a quantum 
meter. We evaluate the probability distributions and the noise 
spectrum of the current.

We acknowledge discussions with Y.~Blanter, L.~Dreher, S.~Gurvitz, D.~Ivanov, 
and A.~Korotkov.

\end{document}